\documentclass{PoS}
\usepackage{subfigure}
\title{
\vspace*{-3cm}
\begin{flushright}\texttt{\footnotesize
CERN-PH-TH/2008-217}
\end{flushright}
\vfill
Revisiting strong coupling QCD \\ at finite temperature and baryon density} 

\ShortTitle{Revisiting strong coupling QCD at finite $T$ and $\mu$} 

\author{\speaker{Michael Fromm}%
	\\
	Institute for Theoretical Physics, ETH Zurich, CH-8093 Zurich, Switzerland\\ 
	E-mail: \email{fromm@phys.ethz.ch}} 
	
\author{Philippe de Forcrand\\ 
Institute for Theoretical Physics, ETH Zurich, CH-8093 Zurich, Switzerland\\
and\\
CERN, Physics Department, TH Unit, CH-1211 Geneva 23, Switzerland\\ 
E-mail: \email{forcrand@phys.ethz.ch}} 
\abstract{
The strong coupling limit ($\beta_{\rm gauge}=0$) of lattice QCD with
staggered fermions enjoys the same non-perturbative properties as continuum
QCD, namely confinement and chiral symmetry breaking.
In contrast to the situation at weak coupling, the sign problem which appears
at finite density can be brought under control for a determination of 
the full $(\mu,T)$ phase diagram by Monte Carlo simulations.
Further difficulties with efficiency and ergodicity of the simulations,
especially at the strongly first-order, low-$T$, finite-$\mu$ transition,
are addressed respectively with a worm algorithm and multicanonical sampling.
Our simulations reveal sizeable corrections to the old results of
Karsch and M\"utter \cite{Karsch:1988zx}. 
Comparison with analytic mean-field determinations of the phase diagram
shows discrepancies of ${\cal O}(10)$ in the location of the QCD critical point.
} 

\FullConference{The XXVI International Symposium on Lattice Field Theory \\ July 14 - 19, 2008\\ Williamsburg, Virginia, USA} 

\begin{document} 
\section{Introduction}
\begin{figure}[t]
\begin{center}
\includegraphics*[scale=.32]{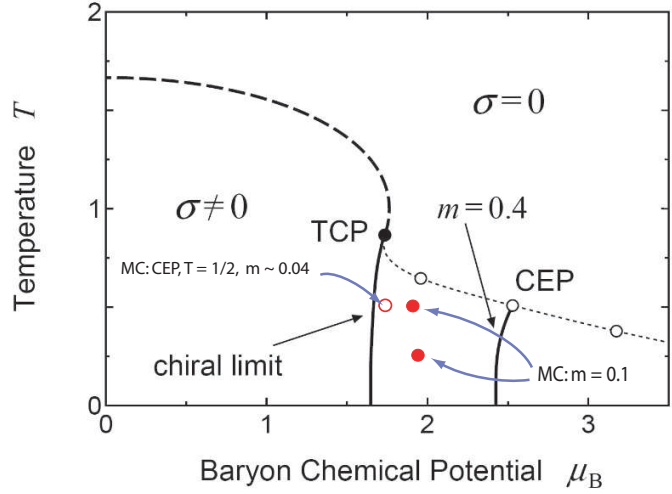}
\label{fig_mf_pd}
\caption{Phase diagram of strong coupling QCD obtained analytically 
in the mean-field approximation, from \cite{Nishida:2003fb}.
For vanishing quark mass, the first-order transition (solid line) at low
temperature turns second-order (dashed line) at a tricritical point (TCP).
For non-zero quark mass, the first-order transition ends in a critical
endpoint (CEP), whose trajectory as a function of the quark mass is shown
by the dotted line. Also displayed are points obtained by the present MC-study: Red dots are transition points for $m=0.1$, the red circle shows the CEP for $T = 1/2$: ($m \approx 0.038, \mu \approx 0.58$).}
\end{center}
\vspace*{-0.5cm}
\end{figure}

We consider lattice QCD with $N_c=3$ colors and 1 species of staggered
fermions (i.e. 4 continuum flavors) at vanishing gauge coupling.
The grand canonical partition function is given by
\begin{equation} 
Z (m, \mu) = \int\mathcal{D}U\mathcal{D}\bar\chi\mathcal{D}\chi\,\mathrm{e}^{S_\mathrm{F}}\label{part_function}
\end{equation}
with
\begin{equation}
	S_\mathrm{F} = \sum_{x,\nu}\bar\chi_x\left[\eta_{x,\hat{\nu}}U_{x,\hat\nu}\chi_{x+\hat\nu} - 	\eta^{-1}_{x,\hat\nu}U^{\dagger}_{x-\hat\nu,\hat\nu}\chi_{x-\hat\nu}\right] + 2m\sum_x\bar\chi_x\chi_x\label{sc_action} \quad ,
\end{equation}
where $\eta_{x,\hat\nu} = \mathrm{e}^{\mu}$ $(\nu = 0)$ and $(-1)^{\sum_{\rho<\nu}x_\rho}$ 
otherwise\footnote{In our notation, all quantities $m, \mu, T,..$ are 
dimensionless, and should be understood as including the appropriate power
of the lattice spacing $a$.}. 
Like continuum QCD, this model shows confinement and chiral symmetry breaking.
Therefore, it has been the object of numerous analytic studies since the
earliest days of lattice QCD, focusing on the mass spectrum 
\cite{Hoek:1981uv, KlubergStern:1982bs} and the $(\mu,T)$ phase diagram
\cite{Damgaard:1985np, Nishida:2003fb, Kawamoto:2005mq}, using increasingly 
refined treatments, all based on the mean-field approximation. 
These investigations are continuing to this day \cite{Miura:2008gd}.
In contrast, very few numerical studies have been performed.
Karsch and M\"utter \cite{Karsch:1988zx} showed how to express $Z(m,\mu)$
as a gas of loops, the monomer-dimer-polymer (MDP) ensemble, where the
sign problem arising at finite chemical potential is very much reduced.
They used this formulation to locate the $T\approx 0$ finite-$\mu$ transition,
which they found to agree with mean-field predictions.
Karsch et al. \cite{Boyd:1991fb} also found the $\mu=0$ finite-$T$ transition 
to be consistent with the expected $O(2)$ universality class in the chiral 
limit, turning into a crossover for finite quark mass.
But a more complete determination of the $(\mu,T)$ phase diagram is lacking.
Moreover, Azcoiti et al. \cite{Aloisio:1999nk} reported ergodicity problems 
with the MDP algorithm of \cite{Karsch:1988zx}, casting some doubt on the
$T\approx 0$, finite $\mu$ results.
This is particularly interesting because of a mismatch between the critical
value of the chemical potential at $T\approx 0$, $\mu_c(T\approx 0, m = 0) = 0.66$,  according to both mean-field \cite{Kawamoto:2005mq} and Monte Carlo \cite{Karsch:1988zx}, and
the baryon mass $M_B \approx 3$ according to mean-field. One would expect
$\mu_c \approx M_B/3$, unless the nuclear interaction is strong.
A determination of $M_B$ was performed in \cite{deForcrand:2006gu}, 
using conventional HMC at $\beta_{\rm gauge} = 0$. 
The value agrees closely with mean-field.
Thus, we now want to determine the $(\mu,T)$ phase diagram by Monte Carlo
simulations, paying special attention to the value of $\mu_c(T\approx 0)$.

The phase diagram can be understood from symmetry considerations.
In the chiral limit $m=0$, the action Eq.~(\ref{sc_action}) enjoys the
staggered $U(1)\times U(1)$ symmetry
\begin{equation}
\chi \rightarrow e^{i\phi_V+i\varepsilon (x) \phi_A}\chi,\quad \bar\chi \rightarrow e^{-i\phi_V+i\varepsilon (x) \phi_A}\bar\chi, \quad \varepsilon(x) = -1^{\sum_{\rho=0}^d x_\rho}\,
\end{equation}
where $U(1)_A$ breaks spontaneously at small $\mu$ and $T$, giving rise
to a chiral condensate $\sigma \equiv -\langle\sum_{a,x}\bar\chi^a_x\chi^a_x\rangle$. Different mean-field approximations all lead to a phase diagram which is qualitatively
similar to Fig.~\ref{fig_mf_pd} with some quantitative differences.
The symmetry is restored at large $T$ or $\mu$, by a phase transition which
is first-order at low temperature, turning second-order at a tricritical
point (TCP), much like what is expected in real QCD with $N_f=2$ massless flavors.
For non-zero quark mass $m$ where $U(1)_A$ is broken explicitly, the second-order transition becomes crossover,
and the TCP becomes a critical endpoint (CEP). The behaviour of the CEP as
the quark mass is increased is of particular interest, given recent 
unexpected findings for $N_f=3$ and $(2+1)$ flavors on coarse lattices
\cite{deForcrand:2006pv}. 
The dotted line in Fig.~\ref{fig_mf_pd} shows the trajectory of 
the CEP in the mean-field approximation: it agrees with conventional 
expectations for real QCD.

\section{Theoretical background}
\begin{figure}[t]
\begin{center}
\subfigure[]{\includegraphics[scale=.25,angle=90]{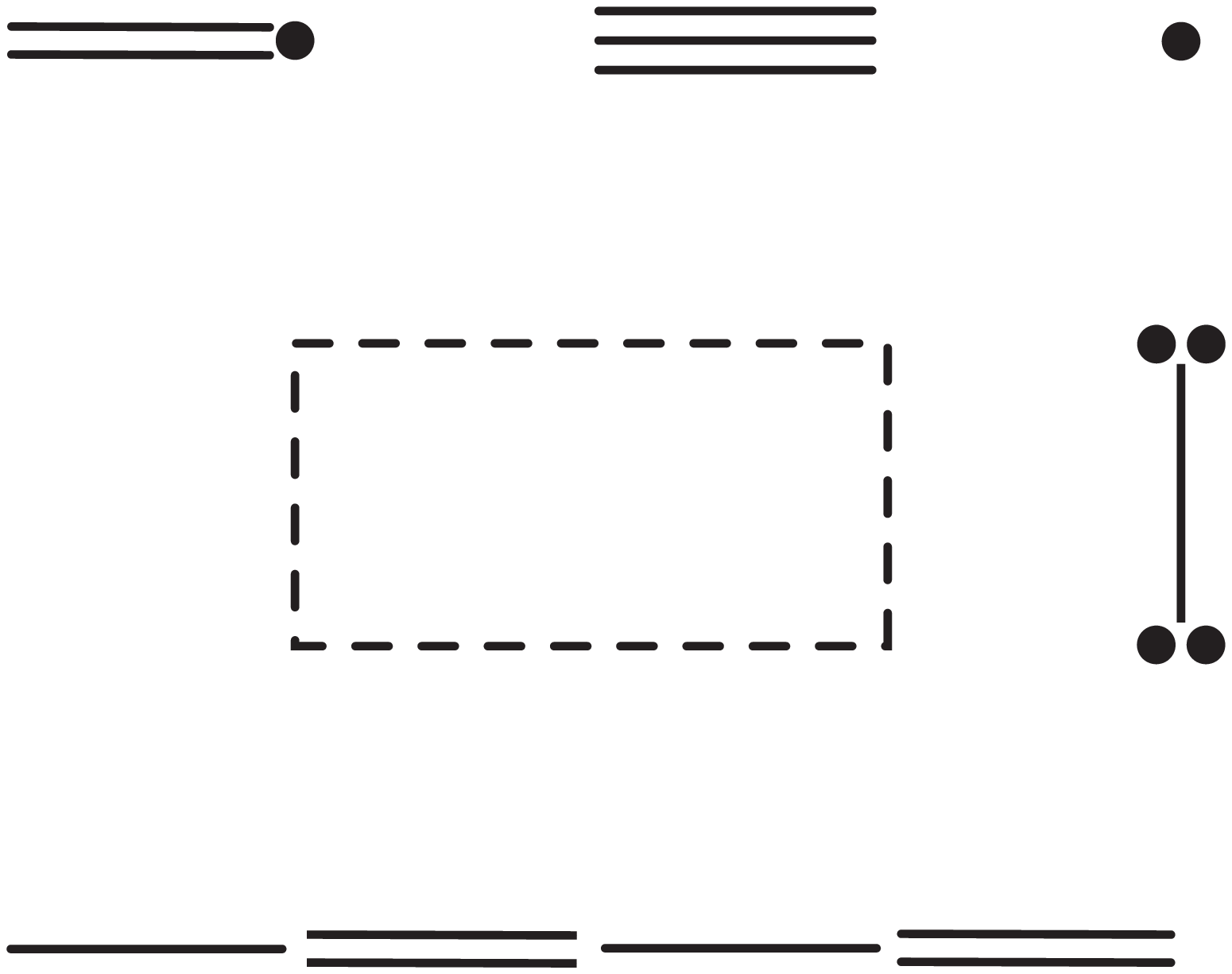}\label{fig_conf_bary}}
\hspace*{1.5cm}
\subfigure[]{\includegraphics[scale=.25, angle = 90]{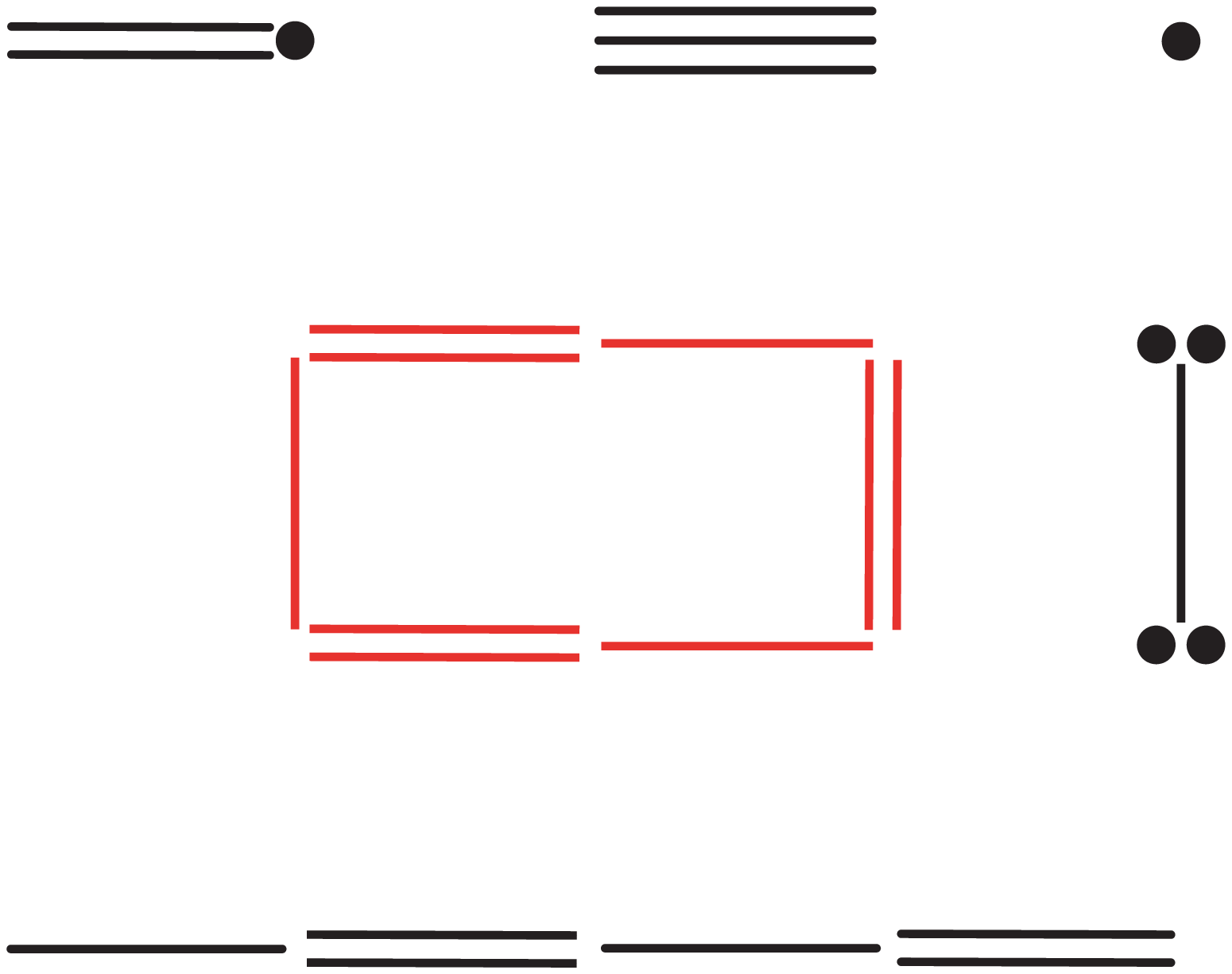}\label{fig_conf_poly}}
\caption{Sample configuration with ($a$) baryonic loop (dashed line) and ($b$) $D_1-D_2$ polymer loop.}
\label{fig_conf}
\end{center}
\vspace*{-0.5cm}
\end{figure}

The usual strategy to deal with the Grassmann fields $\chi, \bar\chi$ 
in Eq.~(\ref{part_function}) is to integrate them out, which yields the customary
fermion determinant. At strong coupling, the absence of a gauge action
allows for an alternative strategy: one integrates over the gauge links
$U$ first \cite{Rossi:1984cv}. This leads to (for $N_c=3$):

\begin{equation}
        Z(m,\mu) = \int\mathcal{D}\bar\chi\mathcal{D}\chi \mathrm{e}^{2m\sum_x\bar\chi_x\chi_x} \prod_{x,\nu} F_{x,x+\hat\nu}\label{part_function_F}
\end{equation}
with
\begin{equation}
        F_{x,x+\hat\nu} = \displaystyle\sum_{k=0}^3\alpha_k(M_xM_{x+\hat\nu})^k + \left[\bar{B}_xB_{x+\hat\nu}\eta_{x,\hat\nu}^3 - \bar{B}_{x+\hat\nu}B_x\eta_{x,\hat\nu}^{-3}\right], \quad \alpha_k = \frac{(N_c-k)!}{N_c!k!}\,.
\end{equation}

The new degrees of freedom are color singlets: monomers 
$M_x = \sum_{a} \bar\chi_{ax}\chi_{ax}$,
dimers $D_{k,xy} = \frac{1}{k!}(M_xM_y)^k, k=1,2,3$,
and baryons and antibaryons 
$B_x = \frac{1}{6} \varepsilon_{abc} \chi_{ax}\chi_{bx}\chi_{cx}$,  
$\bar B_x = \frac{1}{6} \varepsilon_{abc} \bar\chi_{cx}\bar\chi_{bx}\bar\chi_{ax}$.
Moreover, the Grassmann integration generates a ``close-packing'' constraint:
exactly $N_c$ quarks and $N_c$ antiquarks must be present at each site.
This implies that baryon loops $C_B$, representing 
$\Pi_{\langle x,y\rangle \in C_B} \bar{B}_xB_y$ in Eq.~(\ref{part_function_F}),
are self-avoiding. It also implies, for the other sites $x$, that
$n_x + \sum_{b_x}{n_{b_x}} = N_c$, where $n_x$ is the number of monomers on $x$
and $n_{b_x}$ the dimer occupation number (bond number) of the link $b_x$ 
connected to $x$ (see Fig.~\ref{fig_conf}). 
Taking this into account we arrive at the final expression
\begin{equation}
        Z(m,\mu) = \displaystyle\sum_{\left\{n_x,n_b,C_B\right\}}\prod_{b}\frac{(N_c-n_b)!}{N_c!n_b!}\prod_x\frac{N_c!}{n_x!}(2m)^{n_x}\prod_{C_B}w(C_B)\label{bary_part_function} \quad .
\label{part_function_sample}
\end{equation}
Baryon loops $C_B$ come in two orientations ($\pm$) and carry weight
\begin{equation}
w(C_B,\pm) = \varepsilon(C_B) \exp(\pm 3\ell L_t \mu) \,,
\end{equation}
where $\varepsilon(C_B)$ is a sign factor depending on the loop geometry,
and $\ell \geq 0$ is the winding number around the time direction of extent $L_t$.
There also exist self-avoiding, non-oriented meson loops, which consist of
alternating $D_1$ and $D_2$ dimers (see Fig.~\ref{fig_conf}(right)).
Thus, for a given loop geometry $C$, the partition function should sum
over 4 types of loops: two baryon loops with weights $w(C,\pm)$, and 
two meson loops related by $D_1 \leftrightarrow D_2$, with weight $+1$.
Karsch and M\"utter \cite{Karsch:1988zx} had the idea of regrouping these 
4 contributions in 2 sets, by associating with each meson loop half of 
$(w(C,+) + w(C,-))$. In this way, only non-oriented polymer loops $C$
enter in the partition function, with weight
\begin{equation}
\left( 1 + \varepsilon(C) \cosh(3\ell L_t \mu) \right) \quad .
\label{loop_weight}
\end{equation}
The sign problem, which would have been severe in Eq.~(\ref{bary_part_function}),
is now much milder. In particular, for $\mu=0$ one recovers non-negative 
weights. Moreover, it turns out that the sign problem is very much reduced
in comparison with that present in the determinant approach, when one
integrates over fermions first.
This makes the study of QCD at large $\mu$ and low $T$ possible.

\section{Algorithmic issues}
\label{sec_algo}
\begin{figure}[t]
\begin{center}
\includegraphics[scale=.55]{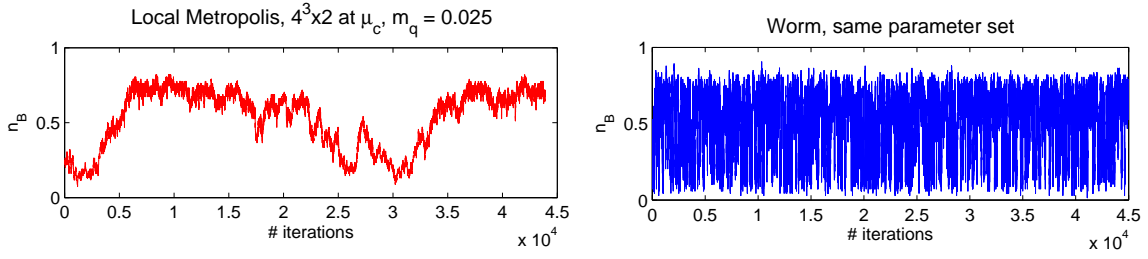}
\vspace*{-0.5cm}
\caption{Monte Carlo history of baryon density, for local Metropolis 
({\em left}) and Metropolis+global worm updates ({\em right}), 
for runs with $m = 0.025, 4^3\times2$ at $\mu_c \approx 0.565$ using similar CPU time.}\label{fig_effi}
\end{center}
\vspace*{-0.5cm}
\end{figure}
The MDP system defined by (\ref{part_function_sample}), (\ref{loop_weight}) was 
sampled in \cite{Karsch:1988zx} using a local Metropolis algorithm which 
operates on pairs of neighbouring sites and tries to replace two monomers 
(one on each site) by a dimer on the connecting link or vice versa. 
Since a monomer carries weight $\sim m$, this prescription is not ergodic 
in the chiral limit or the infinite-mass limit. Simulations in \cite{Karsch:1988zx}
were performed over a narrow range of masses. Even so, ergodicity problems 
were later reported by Azcoiti et al.\cite{Aloisio:1999nk}, 
which cast some doubt on the results of \cite{Karsch:1988zx}. 
Therefore we supplement the local Metropolis update above by a worm algorithm 
\cite{Prokof'ev:2001zz}, which was first adapted to strongly coupled gauge 
theories in \cite{Adams:2003cca}. In Fig.~\ref{fig_effi} we show the Monte Carlo
history of the baryon density in a $4^3\times 2$ system at low quark mass $m$ 
at the critical $\mu$, for both algorithms. The computer time spent is similar in
both cases. In the Metropolis case, changing the baryon density proceeds via
changing the monomer density, which is very unlikely. In the worm case,
a pair of monomers (the head and tail of the worm) is created; then the head
is propagated in a succession of nearest-neighbor hops until it meets again
the tail and annihilates with it, yielding a new contribution to $Z$ having
the same monomer density, but substantially different baryon density.
This allows for efficient simulations over the complete range of quark masses.
In particular, simulating near or at the chiral limit poses no special problem.

Nevertheless, we still have another difficulty: at low temperature, the
finite-$\mu$ transition becomes strongly first-order, which makes a correct
sampling of the low- and high-density phases problematic.
To address this issue, we employ Wang-Landau sampling~\cite{WLR}
to extract an estimator for the probability 
\begin{equation}
P(O,\mu,m)\sim \sum_{k = \{n_x,n_b,C\}}{\delta\left(O-O(k)\right) w_k}
\end{equation}
of a suitable observable $O$ such as the energy density 
$\epsilon = \frac{1}{V}\frac{\partial}{\partial T} \log{Z}$ 
or baryon density 
$n_B = \frac{1}{3V}\frac{\partial}{\partial \mu} \log{Z}$. 
The inverse of the resulting histogram is then used as a weight for 
multicanonical simulations.
Note that the weight $w_k$ defined in Eqs.~(\ref{part_function_sample}) and 
(\ref{loop_weight}) allows for reweighting in both, $\mu$ and $m$, 
once the numbers of monomers and loops with winding number $\ell$ 
for each configuration are known. Hence, the resulting data from multicanonical 
sampling can be safely reweighted to parameter regions where the corresponding 
histogram is sufficiently flat.
\begin{figure}[t]
\subfigure[]{\includegraphics[scale=.35]{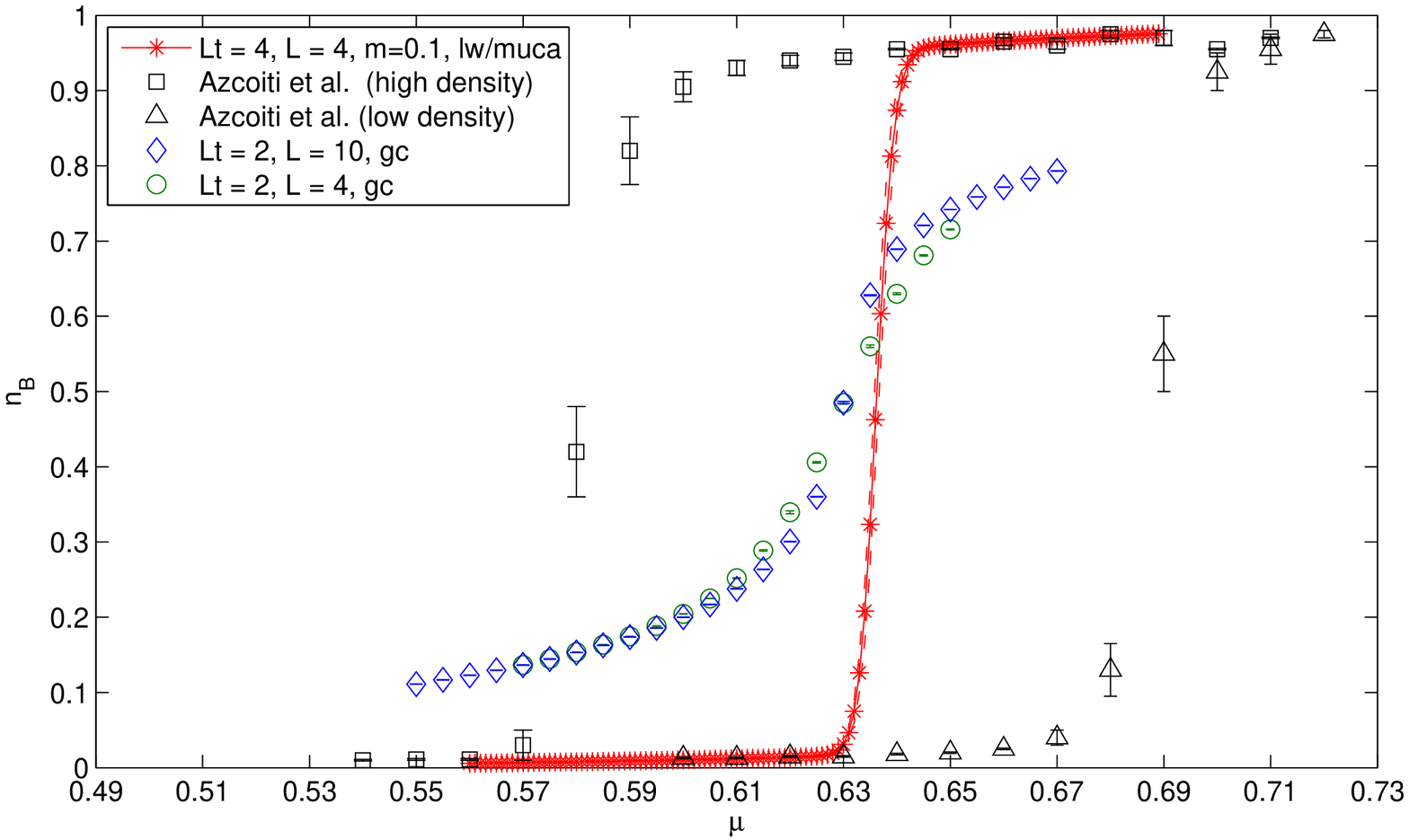}\label{n_B_m01}}
\subfigure[]{\includegraphics[scale=.35]{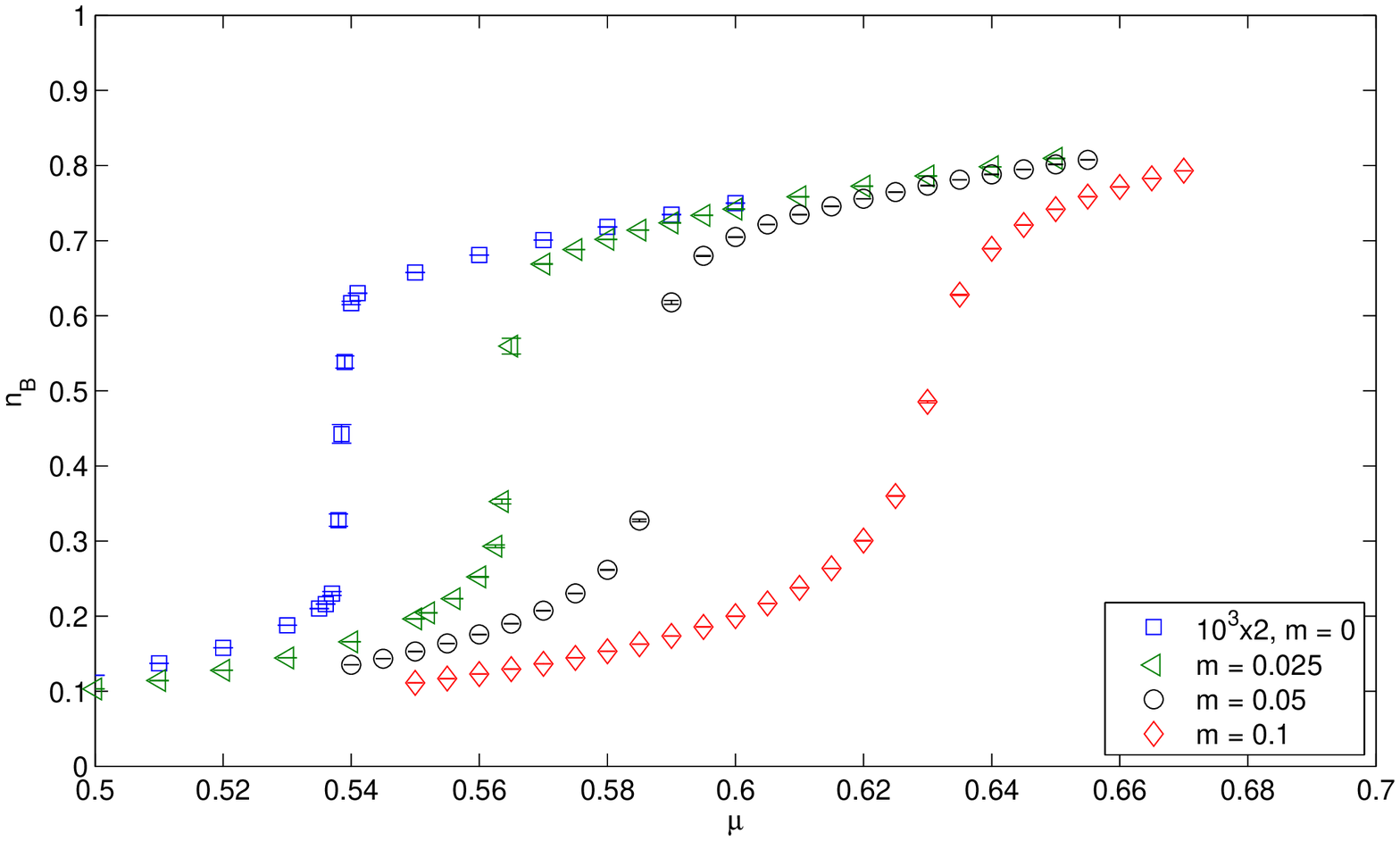}\label{fig_n_B_m_varying}}
\caption{($a$) Baryon density $n_B$ versus chemical potential, 
for systems $L^3\times L_t$ at $m=0.1$, 
($b$) same, for masses $m = 0, 0.025, 0.05, 0.1$ and system $10^3\times 2$.}
\vspace*{-0.5cm}
\end{figure}

\section{Numerical results}
\label{sec_res}
\begin{figure}[t]
\begin{center}
\includegraphics[scale=.55]{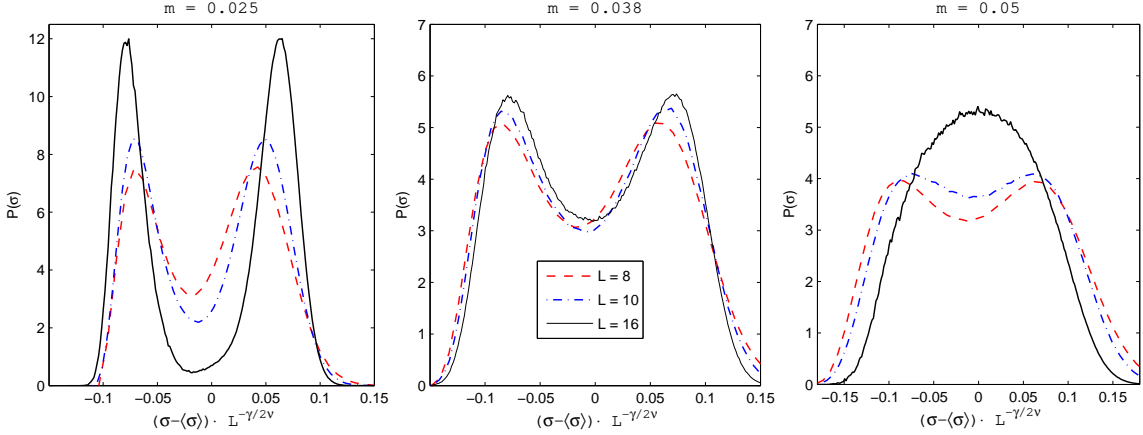}
\vspace*{-0.5cm}
\caption{Probability distribution $P(\sigma)$  for masses $m = 0.025, 0.038, 0.05$ 
tuned to criticality on systems $L^3\times2$, $L = 8,10,16$.
A satisfactory data collapse is observed for the middle mass, using
Ising critical exponents.}
\label{fig_hist_array}
\end{center}
\vspace*{-.5cm}
\end{figure}
We first reproduced the $\mu=0, T\approx 0$ results of 
\cite{deForcrand:2006gu} for the chiral condensate, meson and baryon masses.
This is a non-trivial consistency check between the two strategies of first
integrating over the gauge links (this work) or the fermions 
(followed by HMC) \cite{deForcrand:2006gu}.

Then, we fix the quark mass to $m=0.1$ and perform a comparison with 
\cite{Karsch:1988zx}, who found a phase transition at $\mu \approx 0.69$ on
an $8^3\times 4$ system, i.e. at temperature $T=1/4$. 
Fig.~\ref{n_B_m01} confirms the problems of ergodicity reported by
Azcoiti et al. when using the local Metropolis 
(data taken from \cite{Aloisio:1999nk}). Already on a $4^4$ system, the simulation remains on one of the two metastable branches (density near zero (black triangles) or near the saturation
value of 1 baryon per site (black squares)), and one is unable to determine the
critical value $\mu_c$ of the chemical potential. 
The Wang-Landau/multicanonical approach allows
an ergodic sampling of the critical region. The results (red stars, continuous line)
indicate $\mu_c \approx 0.64$, which is significantly smaller than the
value $\approx 0.69$ of~\cite{Karsch:1988zx}, presumably obtained from the end of the metastability branch.
Fig.~\ref{n_B_m01} also shows higher temperature ($L_t=2$) results
indicating a smoother transition (as expected) and a slight shift of
$\mu_c$ to smaller values (as opposed to the theoretical predictions of Fig.~\ref{fig_mf_pd}).

We now take full advantage of the worm algorithm and explore the chiral
limit at fixed temperature $T=1/2$ in Fig.~\ref{fig_n_B_m_varying}.
As the quark mass is reduced, $\mu_c$ shifts to smaller values, 
and the transition becomes first-order, in qualitative agreement 
with expectations. The order of the transition can only be ascertained
by a finite-size scaling study, which is illustrated in Fig.~\ref{fig_hist_array}.
For increasing mass values, the 3 panels each show the probability distribution of the chiral condensate
for 3 spatial volumes.
In each case, the distribution is reweighted to the pseudo-critical value of $\mu$.
The transition is clearly first-order for the smallest mass, and crossover
for the largest. 
For the middle mass, an approximate data collapse is obtained by rescaling
the condensate by $L^{\frac{\gamma}{2\nu}}$, using $\gamma = 1.237$ and 
$\nu = 0.631$ characteristic of the $3$d Ising universality class.
Indeed, for $m>0$ the U(1) chiral symmetry is broken explicitly, 
so we expect no special symmetry breaking other than $Z(2)$ to happen
at the transition.
Thus, we determine the critical endpoint for temperature $T=1/2$ to be
approximately $(m_c\approx 0.038, \mu_c\approx 0.58)$ (see Fig.\ref{fig_mf_pd}). 
This can be compared with the mean-field prediction 
$(m_c\approx 0.4, \mu_c\approx 0.81)$ (Fig.~\ref{fig_mf_pd}) \cite{Nishida:2003fb}.
The discrepancy is an order of magnitude in $m_c$!

\section{Outlook and conclusions}
We have presented first results of our study of
the strong coupling limit of lattice QCD at finite $\mu$ and $T$.
The comparison with mean-field results and with the original Monte Carlo study 
of Karsch et al. clearly justifies our project. 
Algorithmic advances largely suppress the ergodicity problems of the
latter study, and lead to reliable, new estimates for $\mu_c$ over a range of masses.
The large discrepancy between exact Monte Carlo and approximate mean-field
determinations of the CEP at $T = 1/2$ emphasizes the need for an exact
determination of the whole phase diagram. Quantitative mean-field results should be
considered with caution. \\
Our next step includes the determination of the phase diagram in the chiral 
limit, and the introduction of asymmetric couplings to vary the temperature 
continuously. As simulations in the chiral limit do not pose any problem for 
a wide range of parameters, further topics of interest might include a detailed
comparison with chiral perturbation theory, and $\rho \rightarrow \pi\pi$ decay.

\section{Acknowledgements}
The work of M.F. was supported by ETH Research Grant TH-07 07-2.

\end{document}